\documentstyle[a4,psfig]{article}
\begin{document}

\title{Temperature-extended Jarzynski relation:\\
Application to the numerical calculation of the surface tension}

\author{C. Chatelain\\
Groupe M, Laboratoire de Physique des Mat\'eriaux,\\
Universit\'e Henri Poincar\'e Nancy I,\\
BP~239, Boulevard des aiguillettes,\\
F-54506 Vand{\oe}uvre l\`es Nancy Cedex, France\\
{\small\tt chatelai@lpm.u-nancy.fr}}
\maketitle

\begin{abstract}
We consider a generalization of the Jarzynski relation to the case where the system
interacts with a bath for which the temperature is not kept constant but can vary
during the transformation. We suggest to use this relation as a replacement
to the thermodynamic perturbation method or the Bennett method for the estimation of
the order-order surface tension by Monte Carlo simulations. To demonstrate the
feasibility of the method, we present some numerical data for the 3D Ising model.
\end{abstract}


\def\build#1_#2^#3{\mathrel{
\mathop{\kern 0pt#1}\limits_{#2}^{#3}}}
\def\ket#1{\left| #1 \right\rangle}
\def\bra#1{\left\langle #1\right|}
\def\braket#1#2{\left\langle\vphantom{#1#2} #1\right.
\left| \vphantom{#1#2}#2\right\rangle}
\def\spring{\hskip 0pt minus 1fil}
\def\identite{{\rm 1}\hbox to 1 pt{\spring\rm l}}
\def\trace{\mathop{\rm Tr}\nolimits}

\section{Introduction}
In the last decade, important progresses have been made in the context of far-from-equilibrium
statistical physics with the derivation of the so-called fluctuation theorems. Among these
theorems, the Jarzynski relation has certainly become the most famous one and is now widely
applied both in experimental and numerical studies not only in physics but in chemistry or
biophysics as well. Remarkably, this relation relates thermodynamic equilibrium
quantities to out-of-equilibrium averages over all possible histories of the system when
submitted to a well-defined protocol. The system is initially prepared at thermal equilibrium
and then driven out-of-equilibrium by varying a control parameter $h$ from say $h_1$ to $h_2$.
The work $W$ is recorded for each experiment. The Jarzynski relation states that
        \begin{equation}
        e^{-\beta\Delta F}=\langle e^{-\beta W}\rangle
        \label{eqJarzynski}
        \end{equation}
where $\Delta F$ is the free-energy difference $\Delta F=F(h_2)-F(h_1)$ between the two
equilibrium states at values $h_1$ and $h_2$ of the control parameter. In his first paper on the
subject, Jarzynski gave a derivation of this relation in the case of a set of particles
evolving according to the laws of Newtonian mechanics~\cite{Jarzynski97a}. The derivation depends
crucially on the assumption that the system is initially thermalized with a heat bath at
temperature $T$ but then isolated from it while the work is exerted. No exchange of heat with the
environment is possible. A change of the temperature of the bath has thus no consequence. 
In a second paper~\cite{Jarzynski97b}, Jarzynski showed that the relation (\ref{eqJarzynski}) may
be established in the case of a Markovian dynamics too. In this case, the interaction with the
bath is encoded in the transition rates. The demonstration does not require anymore the insulation
of the system from the bath while the work is exerted. Exchange of heat with the environment is
properly taken into account during the whole process.
\\

It is a simple exercise to generalize the Jarzynski relation to the case where the
temperature of the heat bath changes with time. In the special case where no work is
exerted on the system, one obtains (\ref{Jarzynski2}):
	\begin{equation}
	e^{-\Delta(\beta F)}=\langle e^{-\int E(t)\dot\beta dt}\rangle
	\label{eqJarzynski3}
	\end{equation}
where $E(t)$ is the total energy of the system at time $t$. The derivation can be found
in the appendix. As noted by Crooks in his dissertation~\cite{CrooksDissert} ``it is possible
to view this change as an additional perturbation that exerts a sort of {\sl entropic work}
on the system''. Like any other Jarzynski relation, Eq. (\ref{eqJarzynski3}) is also
a special case of the {\sl annealed importance sampling} method~\cite{Neal01}.
As far as we know, the case of a varying temperature has not attracted any attention.
The aim of this paper is to show that a Jarzynski relation generalized to a temperature change
may also have some interesting applications and deserves interest. As an example, we propose a
new way to estimate by Monte Carlo simulations the order-order surface tension. This
method is tested for the three-dimensional Ising model.
The surface tension is related to the remaining free energy of a system containing
two phases in coexistence separated by an interface when the contribution of these bulk phases
has been removed. The Jarzynski relation offers the possibility to calculate these free energies
in a Monte Carlo simulation.
\\ 

The precise numerical determination of the surface tension has attracted a continuous interest
for several decades. It is indeed an important quantity in nucleation theory and may be used
to distinguish between a first-order phase transition and a continuous one. In the latter case,
the surface tension is expected to vanish at the critical point. In the context of lattice spin
models, the interface tension permitted to show for example the existence of a
randomness-induced continuous transition in the 2D eight-state Potts model~\cite{Chen95}
or that the phase transition of the 3D 3-state Potts model is weakly first-order~\cite{Janke96}.
In chemistry, one is interested in the surface tension in fluids or colloids and the 3D Ising
model is commonly used as a toy model for the description of the liquid-vapor phase
transition~\cite{Provata97,Jain03}. In the context of high-energy physics, the
confinement-deconfinement phase transition between hadronic matter and quark-gluon plasma
is a very active research field. Numerous Monte Carlo simulations have been devoted to the
estimation of the interface tension of the pure SU(3) gauge model~\cite{Huang90,Grossmann92,
Alves92,Brower92,Hackel92,Papa94}. The Ising model is also of interest in this context since
duality maps the 3D $Z_2$ Yang-Mills gauge theory onto the 3D Ising model~\cite{Caselle06}.
Wilson loop and Polyakov loop correlators are then related to interfaces in the Ising model. 
In all theses situations where the surface tension is of interest, numerical accuracy has
been constantly improved not only by the exponential growth of computer speed but mainly by
progresses in algorithms (multicanonical simulations~\cite{Berg92}, flat histogram
methods~\cite{Jain03},$\ldots$) and in the protocol set up to have access to the surface
tension (thermodynamic integration, Binder's histogram method~\cite{Binder82},
snake algorithm~\cite{Pepe01,Pepe02,Forcrand05,Forcrand05b}, $\ldots$).
\\

This paper is organized as follows: in the first section, we expose the method used to
estimate numerically the order-order surface tension. The derivation of the Jarzynski
relation is given in the appendix. Numerical results for the 3D Ising model are presented
in the second section. Conclusions and discussion follow.

\section{Numerical estimation of the surface tension}
Consider the classical ferromagnetic Ising model defined by the usual Hamiltonian~\cite{Ising25}
        \begin{equation}
	{\cal H}_{\rm Ising}=-J\sum_{(i,j)} \sigma_i\sigma_j,
        \quad \sigma_i=\pm 1
	\end{equation}
where the sum extends over nearest neighbors only. In order to favor the appearance of an
interface in the system, we first impose anti-periodic boundary conditions in one direction.
The free energy $F(T)$ of the system at a temperature $T<T_c$ may be decomposed as
	\begin{equation}
	F(T)=F_{\rm Ferro}(T)+F_s(T)-k_BT\ln L
	\end{equation}
where $F_{\rm Ferro}$ is the contribution of the two ferromagnetic phases on both
sides of the interface, $F_s$ the free energy of the interface and $k_B\ln L$ the
entropy associated to the degeneracy of the position of the interface. One can restrict
the analysis to the case where only one interface appears in the system. Indeed, the
probability for $n$ interfaces behaving as $e^{-\beta n\sigma L^{d-1}}$, the contribution
of spin configurations with more than one interface can be neglected for sufficiently large
lattice sizes $L$. The order-order surface tension is defined as
	\begin{equation}
	\sigma(T)=\lim_{L\rightarrow +\infty} {F_s(T)\over L^{d-1}}.
	\label{DefSigma}
	\end{equation}
To first calculate the free energy $F(T)$, we will estimate numerically the difference
$\Delta(\beta F)={F(T)\over k_BT}-{F(T_i)\over k_BT_i}$. The temperature $T_i\ll T_c$ is
chosen such that the thermal contribution to the free energy $F(T_i)$ can be neglected,
i.e. $F(T_i)=E_0-k_BT\ln 2-k_BT\ln L$ where $E_0=-J(dL^d-2L^{d-1})$ is the energy of
the spin configuration with a flat interface, $k_B\ln 2$ the entropy associated
to the degeneracy of the ground state corresponding to $\sigma_i\rightarrow -\sigma_i$
and $k_B\ln L$ the entropy associated to the position of the interface.
\\

The usual method to compute $\Delta(\beta F)$ is based on the decomposition
	\begin{equation}
	e^{-\Delta(\beta F)}
	={{\cal Z}_{\beta}\over{\cal Z}_{\beta_i}}
	={{\cal Z}_{\beta_{N+1}}\over{\cal Z}_{\beta_N}}
	{{\cal Z}_{\beta_N}\over{\cal Z}_{\beta_{N-1}}}
	\ldots{{\cal Z}_{\beta_2}\over{\cal Z}_{\beta_1}}
	\end{equation}
where $\{\beta_1,\ldots,\beta_{N+1}\}$ is a set of inverse temperatures interpolating
between $\beta_i=\beta_1$ and $\beta=\beta_{N+1}$. When $\beta_k$ and $\beta_{k+1}$ are
sufficiently close, the ratio ${\cal Z}_{\beta_{k+1}}/{\cal Z}_{\beta_k}$ can be estimated
during a single Monte Carlo simulation at the inverse temperature $\beta_k$ by thermodynamic
perturbation:
	\begin{equation}
	{{\cal Z}_{\beta_{k+1}}\over{\cal Z}_{\beta_k}}
	=\langle e^{-(\beta_{k+1}-\beta_k)E}\rangle_{\beta_k}
	\label{PertubThermo}
	\end{equation}
or by more elaborated procedures like the Bennett method~\cite{Bennett76}.
The estimation of $\Delta(\beta F)$ requires $N$ Monte Carlo simulations at the
inverse temperatures $\beta_1,\ldots\beta_N$ with for each of them a
sufficiently large number $n_{\rm exp.}$ of Monte Carlo steps in order to get
accurate averages.
\\

The Jarzynski relation (\ref{Jarzynskib}) offers the possibility to estimate
the quantities ${\cal Z}_{\beta_{k+1}}/{\cal Z}_{\beta_k}$. The system is initially
thermalized at the temperature $T_k=1/k_B\beta_k$. First the initial spin
configuration is stored. Then $n_{\rm iter.}$ MCS are performed with a
temperature increasing from $T_k$ to $T_{k+1}$. The quantity
$e^{-{\beta_{k+1}-\beta_k\over n_{\rm iter.}}\sum_{t=0}^{n_{\rm iter.}
-1}E(t)}$ is calculated and accumulated. The initial spin configuration is
restored and a few additional MCS at temperature $T_k$ are performed to
generate a new initial spin configuration uncorrelated with the previous one.
The whole procedure is repeated $n_{\rm exp.}$ times. Finally the average
$\langle e^{-{\beta_{k+1}-\beta_k\over n_{\rm iter.}}\sum_{t=0}^{n_{\rm iter.}-1}
E(t)}\rangle$ gives $e^{-\Delta(\beta F)}$ where $\Delta(\beta F)=\beta_{k+1}
F(\beta_{k+1})-\beta_k F(\beta_k)$ according to the Jarzynski relation
(\ref{Jarzynskib}).
\\

Both methods lead to a numerical estimate of $F(T)$. The last step is now to calculate
$F_{\rm Ferro}(T)$, i.e. the contribution of the two ferromagnetic phases.
To that purpose, we make a second Monte Carlo simulation with periodic boundary
conditions. The method presented above is applied again to estimate the difference
$\Delta(\beta F_{\rm Ferro})={F_{\rm Ferro}(T)\over k_BT}-{F_{\rm Ferro}(T_i)\over k_BT_i}$.
The approximation $F_{\rm Ferro}(T_i)=E_0-k_BT\ln 2$ where $E_0=-JdL^d$ gives finally
$F_{\rm Ferro}(T)$. We can now use (\ref{DefSigma}) to estimate the surface tension.
\\

The algorithm based on the Jarzynski relation depends on three parameters: $N$ the
number of temperature intervals in which $[\beta_i;\beta]$ is divided, $n_{\rm iter.}$
the number of MCS bringing the temperature from $T_k$ to $T_{k+1}$ and $n_{\rm exp.}$
the number of measurements of $e^{-\Delta\beta\sum E}$ performed. The thermodynamic
perturbation method (\ref{PertubThermo}) corresponds to the special case $n_{\rm iter.}=1$.
The use of the Jarzynski relation may improve the convergence and save computer time
by an optimal choice of the set $(N,n_{\rm iter.},n_{\rm exp.})$ of parameters.
Increasing $n_{\rm iter.}$ allows to decrease $N$ and to some extend $n_{\rm exp.}$.
However, we have no recipe to determine these optimal parameters and moreover, they
probably depend a lot on the model and on the Monte Carlo dynamics chosen.

\section{Application to the 3D Ising model}
The method presented above is applied to the 3D Ising model. We considered the ``extreme''
case $N=1$, i.e. the temperature is not increased step by step in different Monte
Carlo simulations but in a single simulation from $T_i$ to $T$. This choice is certainly
not the optimal one but it demonstrates that the Jarzynski relation can be applied
even when the temperature is changed by a very large amount.
\\

In one of the three directions, the boundary conditions are first anti-periodic
and then periodic. In the two other directions, we are free to choose any boundary
conditions. We considered both periodic (PBC) and free boundary conditions (FBC).
This choice is motivated by the fact that only periodic boundary conditions
are usually considered in the literature but in the 2D Ising model, one can
easily check that finite-size corrections to the surface tension are smaller
with FBC. This can be understood by considering the equivalent one-dimensional
Solid-on-Solid. With FBC, the steps made by the interface are
independent and the free-energy may be written as the sum:
	\begin{eqnarray}
	F_s&=&-k_BTL\ln\Big[\sum_{n=-\infty}^{+\infty} e^{-2\beta J(1+|n|)}\Big]
	\nonumber\\ &=&2JL+k_BTL\ln\tanh\beta J
	\label{ResSOS2D}
	\end{eqnarray}
which leads to Onsager's exact result $\sigma=2J+k_BT\ln\tanh\beta J$ for any
lattice width $L$. When the lattice size perpendicular to the interface is finite,
corrections arise because the sum of (\ref{ResSOS2D}) becomes bounded. PBC impose a
non-local condition that results into additional finite-size corrections.
\\

We considered  cubic lattices with sizes $L=4,6,8,12,16,24,32,48$. The initial
inverse temperature is $\beta_i=2$, to be compared with the critical inverse
temperature $\beta_c\simeq 0.22165\ldots$~\cite{Blote99}. The spins are initialized
in the state $\sigma_i=+1$ if they lie in the upper half of the system and in the
state $\sigma_i=-1$ otherwise, i.e. we start with a flat interface separating two
ferromagnetic phases at saturation. We made $500\times L$ MCS to equilibrate the system
at this temperature. This choice is a very safe bet since in practise very few spin
flips occur during these iterations~\footnote{The probability that a spin be flipped
during a Monte Carlo step is $e^{-4d\beta_iJ}$ where $d$ is the dimension of the lattice.
At the inverse temperature $\beta_iJ=2$ and dimension $d=3$, this probability is as
small as $10^{-14}$.}. The inverse temperature
$\beta$ is then decreased up to the final value in the range $[0.24;0.80]$.
The final temperature is reached after $n_{\rm iter.}=250\times L$ MCS. The experiment
is repeated $n_{\rm exp.}=1000$ times. Before each experiment, the spin configuration
is stored. After the experiment, it is restored and $25\times L$ additional MCS are
performed to generate a new non-correlated spin configuration for the next experiment.
The statistical error on the average $\langle e^{-\sum_t E(t)\Delta_t\beta}\rangle$
is estimated as the standard deviation as expected from the central limit theorem.
However systematic deviations may occur due to the fact that the average is
dominated by rare events. This effect is important when the variation of the
temperature is fast, i.e. when the transformation is strongly irreversible.
Moreover, the bias induced is difficult to quantify. As shown by Gore
{\sl et al.}~\cite{Gore03,Jarzynski05},
``{\sl the number of realizations for convergence grows
exponentially in the average dissipated work}''.
To overcome this difficulty, we took advantage of the fact that the detailed
balance condition (\ref{DetailledBalance}) is the only assumption made on the transition
rates in the derivation of the Jarzynski relation. Any Monte Carlo algorithm satisfying
this condition can thus be used. To make the transformation as reversible as possible,
we used cluster algorithms: the Hasenbush-Meyer algorithm~\cite{Hasenbusch91} for
anti-periodic boundary conditions and the Swendsen-Wang algorithm~\cite{Swendsen87}
for periodic boundary conditions.

\begin{center}
\begin{figure}[!ht]
        \centerline{\psfig{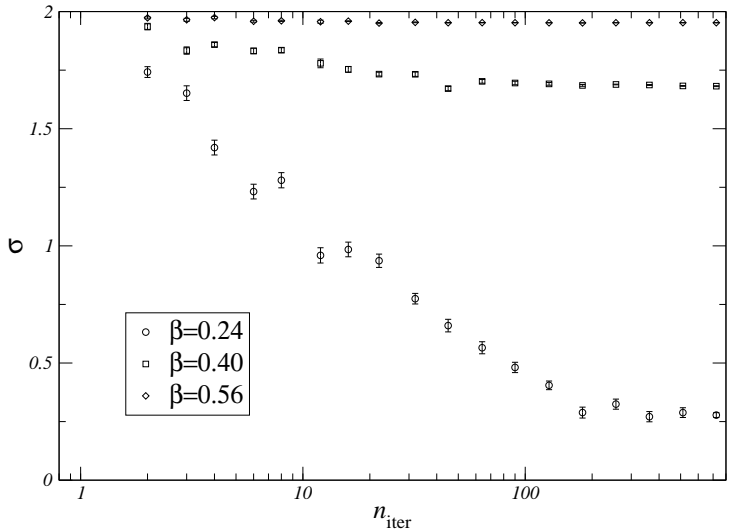}}
        \caption{Surface tension $\sigma$ versus the number of iterations $n_{\rm iter.}$
	for three different temperatures: $\beta=0.24$, $\beta=0.40$ and $\beta=0.56$
	from bottom to top. The lattice size is $L=16$. Error bars on the data correspond
 	to statistical errors.}
        \label{fig1b}
\end{figure}
\end{center}

The surface tension is calculated according to the method presented in section~2.
Estimates of the surface tension for an increasing number of iterations $n_{\rm iter.}$ are
plotted on figure \ref{fig1b}. The number of experiments is kept fixed to the value
$n_{\rm exp.}=1000$ which means that rare events with a probability smaller than
$1/n_{\rm exp.}=10^{-3}$ are not correctly sampled. The figure shows two regimes. When the
number of iterations is large $n_{\rm niter}>n_{\rm iter}^{\rm opt.}$, the estimates display
a plateau. The amplitude of the fluctuations around the value of this plateau is of the same
order of magnitude than the error bars.  In contradistinction, the estimates display
a systematic deviation when the number of iterations is small $n_{\rm niter}
<n_{\rm iter}^{\rm opt.}$. The data indicate a deviation proportional to $\ln{1\over
n_{\rm niter}}$. The contribution of the rare events are larger than the statistical errors
and cannot be neglected. The cross-over between the two regimes depends on the temperature.
Surprisingly, it does not depend much on the lattice size.
\\

We have also tried to estimate the surface tension using the width $\sigma_{W_S}^2
=\langle W_S^2\rangle -\langle W_S\rangle^2$ of the probability distribution of the entropic
work. These distributions are indeed close to Gaussian distributions so the Jarzynski
relation reads
        \begin{equation}
        \langle e^{-W_S}\rangle={1\over\sqrt{2\pi\sigma_{W_S}^2}}
        \int e^{-W_S-(W_S-\langle W_S\rangle)^2/2\sigma_{W_S}^2}dW_S
        \Leftrightarrow\ \Delta(\beta F)=\langle W_S\rangle-{\sigma_{W_S}^2\over 2}
        \end{equation}
The estimates of the surface tension are compatible with those obtained using
the Jarzynski relation $(\ref{eqJarzynski3})$ but the error bars are larger. As a
consequence, we will not consider this estimate in the following.
\\

\begin{center}
\begin{figure}[!ht]
        \centerline{\psfig{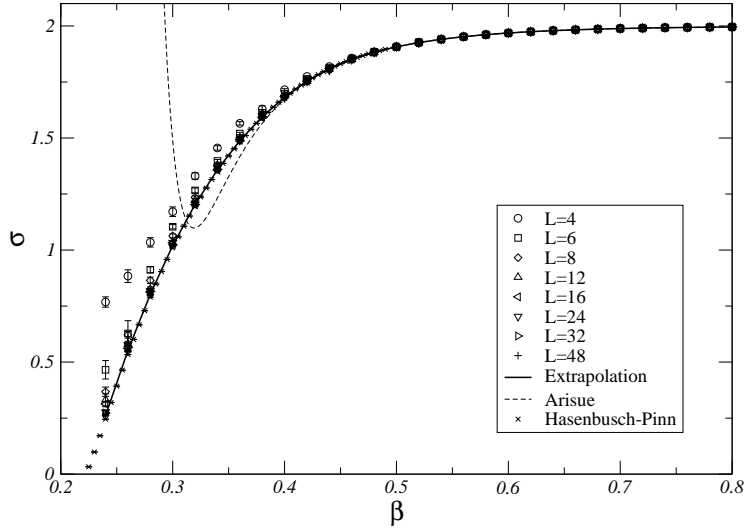}}
        \caption{Numerical estimates of the surface tension of the 3D Ising model with respect
	to the inverse temperature $\beta$ for periodic boundary conditions.}
        \label{fig2}
\end{figure}
\end{center}

\begin{center}
\begin{figure}[!ht]
        \centerline{\psfig{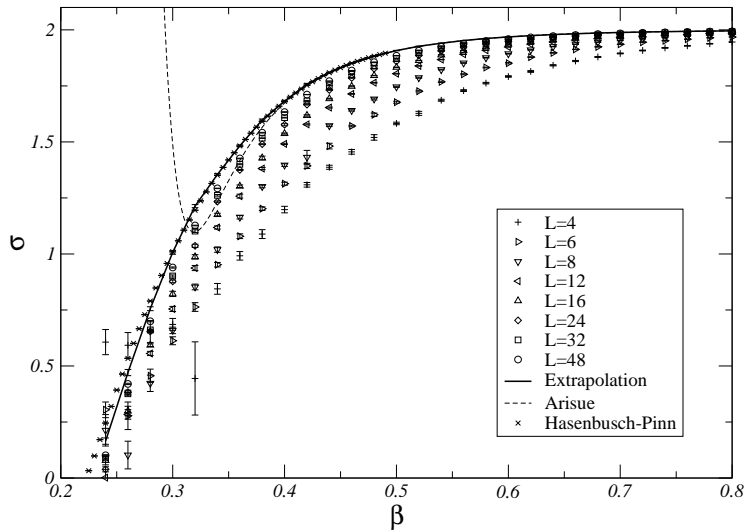}}
        \caption{Numerical estimates of the surface tension of the 3D Ising model with respect
	to the inverse temperature $\beta$ for free boundary conditions.}
        \label{fig3}
\end{figure}
\end{center}

\begin{center}
\begin{figure}[!ht]
        \centerline{\psfig{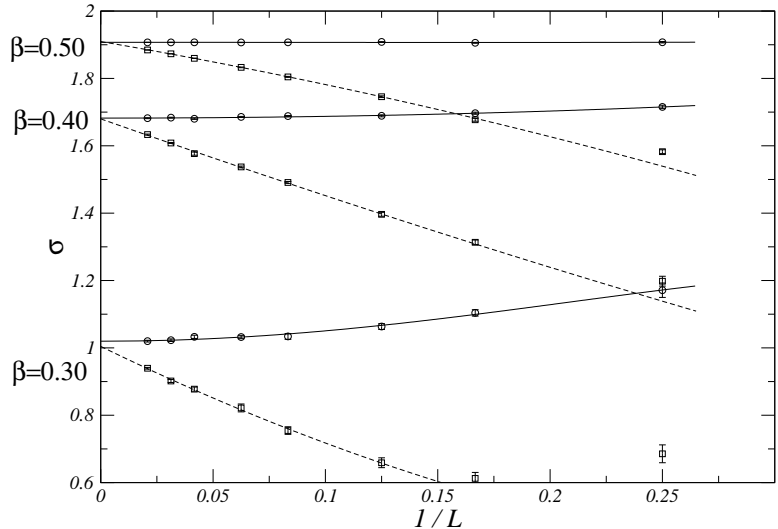}}
        \caption{Extrapolation of the surface tension for three different temperatures.
	The solid curves correspond to periodic boundary conditions and the dashed ones
	to free boundary conditions.}
        \label{fig4}
\end{figure}
\end{center}

Our numerical data of the surface tension are presented on figures~\ref{fig2} (PBC)
and \ref{fig3} (FBC). The few points with large error
bars on figure \ref{fig3} correspond to small systems at temperatures higher than the
finite-size ``critical'' temperature $T_c(L)$. The system is thus already in the paramagnetic
phase and the interface disappears. No order-order surface tension can then be given.
In contradistinction to the two-dimensional case, the surface tension displays larger
finite-size corrections with FBC. The data have been extrapolated to the thermodynamic
limit using the {\sl ansatz}:
	\begin{equation}
	\sigma(L)=\sigma(\infty)+{c_1\over L^2}+{c_2\over L^4}
	\label{FitPBC}
	\end{equation}
for PBC and
	\begin{equation}
	\sigma(L)=\sigma(\infty)+{d_1\over L}+{d_2\over L^2}
	\label{FitFBC}
	\end{equation}
for FBC. The error bars on the value $\sigma(\infty)$ take into account both the errors
on the data and due to the fit. The figure \ref{fig4} gives an example of this extrapolation
for three different temperatures. While the whole range of lattice sizes $L$ are well
reproduced by the ansatz (\ref {FitPBC}) for PBC, the two smallest lattice sizes
($L=4$ and $L=6$) cannot be taken into account by (\ref {FitFBC}) for FBC and have been
discarded. The extrapolated value $\sigma(\infty)$ of the surface tension can be compared
in figures (\ref{fig2}) and (\ref{fig3}) to the 17th-order low-temperature expansion
(see \cite{Hasenbusch93} for the coefficients) and to large-scale Monte Carlo
simulations~\cite{Hasenbusch93}. Our data are perfectly compatible to the latter apart
from a small disagreement for FBC at temperatures close to the critical point.
The too simple ansatz (\ref {FitFBC}) may be responsible for this disagreement.

\section{Discussion and conclusions}
We have discussed a Jarzynski relation generalized to temperature changes. Its derivation
is limited to Markovian dynamics since the interaction with the bath is properly taken
into account only in this case. We have then presented a modified version of the usual
algorithm allowing for the computation of the surface tension by Monte Carlo simulations
where the Jarzynski relation replaces thermodynamic perturbation. The approach
has been tested in the case of the 3D Ising model. Despite the common belief that rare
events prevent the application of Jarzynski relation, the method is efficient because
the transformation does not bring the system far from equilibrium. This is possible
because the method uses initial and final states not very different. Basically,
the final state display thermal fluctuations while the initial state is almost the ground
state. But in both cases, two ferromagnetic phases are separated by an interface. In
contradistinction to other algorithms, the interface is not created during the
transformation. Moreover, we use cluster algorithms. Their relaxation times are
much smaller than the usual Metropolis algorithm and allows the final state to be close to
equilibrium. The systematic deviation due to an insufficient sampling of the rare events
that dominates the average $\langle e^{-W_S} \rangle$ has been shown to be negligible
compared to statistical errors for a reasonable number of iterations. Our aim was only to
prove the usefulness of the Jarzynski relation despite the fact that rare events play an
important role and not to add a 6th or 7th new digit to the estimates of the surface
tension that can be found in the literature. The computational effort we devoted to
the calculation is much smaller (around 100 times) than that of ref.~\cite{Hasenbusch93}
for instance. For $\beta=0.45$ and $L=32$, our estimate of the surface tension is $1.8316(4)$
while Hasenbusch {\sl et al.} obtained a much accurate estimate: $1.831368(5)$. However,
they only calculated numerically the difference $F(\beta=0.45)-F(\beta_i=0.515)$ and
then used the 17th-order low-temperature expansion estimate of $F(\beta=0.515)$ to obtain
the surface tension. We calculated numerically the difference $F(\beta=0.45)
-F(\beta_i=2)$ and estimated $F(\beta_i=2)$ as the $T=0$ ground-state free energy.
By starting with a much temperature, we do not need the 17th-order low-temperature
expansion to calculate the surface tension.
\\

The main drawback of our approach is that the convergence of estimates of the free
energy obtained by the Jarzynski relation is not well understood yet. We cannot
give any recipe to adjust the three parameters $N,n_{\rm iter.},n_{\rm exp.}$
to obtain the best convergence. Research efforts in this direction are highly
desirable.
\\

One can imagine other protocols based on the Jarzynski relation to estimate the surface
tension. One can start first with a strong homogeneous magnetic field on the up
and down boundaries to force the system in the ferromagnetic state. Then one of these
magnetic fields is reversed to favor the appearance of the interface. By recording the
work while the magnetic field is reversed, the free energy could be estimated using
the more usual Jarzynski relation (\ref{eqJarzynski}). However, the convergence may
not be as good because an interface has to be created during the magnetic
field reversal. In the approach based on variation of the temperature, this
interface is already present in the initial state.
\\

The relation (\ref{Jarzynski}) may be useful even in the case of the measurement of
the difference of the free energies of two equilibrium states at the same temperature.
If these two states are separated by a free energy barrier, the relation (\ref{Jarzynski})
allows first to increase the temperature so that the system can pass the barrier
and then to decrease the temperature down to its initial value.

\section*{Acknowledgements}
The laboratoire de Physique des Mat\'eriaux (LPM) is Unit\'e Mixte de Recherche CNRS
number 7556. The author would like to thank Martin Hasenbsuch for having explained
him the variant of the snake algorithm used in reference \cite{Caselle06}.
He also thanks the Statistical Physics group of the LPM for interesting
discussions on this subject and many other ones.

\section*{Appendix: derivation of the Jarzynski relation}
We shall derive the Jarzynski relation along the lines of the derivation given in the
appendix of \cite{Jarzynski97b}. Let us denote $\vec\sigma$ the microscopic states of the
system, i.e. the spin configurations in the case of the Ising model. We shall
assume that the dynamics of the system is Markovian and is governed by the master equation
        \begin{equation}
        \wp(\vec\sigma,t+1)=\sum_{\vec\sigma'} \wp(\vec\sigma',t)
	W_{\beta(t),h(t)}(\vec\sigma'\rightarrow \vec\sigma)
        \label{eq1}
	\end{equation}
where  $\wp(\vec\sigma,t)$ is the probability to find the system in the state $\vec\sigma$
at time $t$ and $W_{\beta(t),h(t)}(\vec\sigma\rightarrow\vec\sigma')$ is the
transition rate per time step from the state $\vec\sigma$ to the state $\vec\sigma'$.
The transition rates take into account the interaction of the system with an
heat bath in an effective manner. They depend on the temperature $T=1/k_B\beta$
of the bath and on an external parameter $h$, for instance the magnetic field.
The master equation is equivalent to the Bayes relation which means that the transition
rates are conditional probabilities: $W_{\beta(t),h(t)}(\vec\sigma\rightarrow\vec\sigma')
=\wp(\vec\sigma',t+1|\vec\sigma,t)$. As a consequence, the transition rates satisfy
the condition 
	\begin{equation}
	\sum_{\vec\sigma'} W_{\beta,h}(\vec\sigma\rightarrow \vec\sigma')=1
	\label{eq1b}
	\end{equation}
When $\beta$ and $h$ are kept constant, the system is expected to evolve towards the
Boltzmann equilibrium distribution
        \begin{equation}
	\pi_{\beta,h}(\vec\sigma)={1\over{\cal Z}_{\beta,h}}
	e^{-\beta(E(\vec\sigma)-hM(\vec\sigma))}
        \label{eq1c}
	\end{equation}
where $M$ is the extensive quantity associated to the control parameter $h$ (for instance
the magnetization when $h$ is the magnetic field). The equilibrium distribution satisfies
the stationarity condition
        \begin{equation}
	\pi_{\beta,h}(\vec\sigma)=\sum_{\vec\sigma'}
	\pi_{\beta,h}(\vec\sigma')W_{\beta,h}(\vec\sigma'\rightarrow \vec\sigma)
        \label{eq2}
	\end{equation}
For convenience one imposes the more restrictive condition of detailed balance
	\begin{equation}
	\pi_{\beta,h}(\vec\sigma)W_{\beta,h}(\vec\sigma\rightarrow \vec\sigma')
	=\pi_{\beta,h}(\vec\sigma')W_{\beta,h}(\vec\sigma'\rightarrow \vec\sigma)
        \label{DetailledBalance}
	\end{equation}
that leads to (\ref{eq2}) when using the condition (\ref{eq1b}).
\\

At time $t=t_i$ the system is thermalized with a bath at the temperature $1/k_B\beta(t_i)$
and external parameter $h(t_i)$, i.e. $\wp(\vec\sigma,t_i)=\pi_{\beta(t_i),h(t_i)}
(\vec\sigma)$. The temperature and the external parameter are then varied and the quantity
 	\begin{equation}
	\langle X(t_i,t_f)\rangle=\langle e^{-\sum_{t=t_i}^{t_f-1} (\Delta_t\beta
	E(\vec\sigma(t))-\Delta_t(\beta h)M(\vec\sigma(t)))}\rangle
	\label{eq4}
	\end{equation}
is measured. The brackets $\langle\ldots\rangle$ denotes an average over all possible
histories of the system between $t_i$ and $t_f$. We have introduced the notations
$\Delta_t\beta=\beta(t+1)-\beta(t)$ and $\Delta_t(\beta h)=\beta(t+1)h(t+1)-\beta(t) h(t)$.
Following Jarzynski~\cite{Jarzynski97b}, each time step consists into two substeps. First
the temperature and the external parameter are changed by an amount $\Delta_t\beta$
and $\Delta_t h$. The latter induces a work $W=-\Delta_t hM$. In the second substep, the
master equation is iterated once so that the system makes a transition to a new state
$\vec\sigma(t+1)$ with the conditional probability $W_{\beta(t+1),h(t+1)}(\vec\sigma(t)
\rightarrow\vec\sigma(t+1))$. The energy change $E(\vec\sigma(t+1))-E(\vec\sigma(t))$
corresponds to the heat exchanged with the heat bath. Rewriting $X(t_i,t_f)$ using the
equilibrium probability (\ref{eq1c}) as
	\begin{equation}
	X(t_i,t_f)=\prod_{t=t_i}^{t_f-1} {{\cal Z}_{\beta(t+1),h(t+1)}
	\pi_{\beta(t+1),h(t+1)}(\vec\sigma(t))
	\over  {\cal Z}_{\beta(t),h(t)}\pi_{\beta(t),h(t)}(\vec\sigma(t))},
	\end{equation}
the average (\ref{eq4}) reads
	\begin{eqnarray}
	\langle X(t_i,t_f)\rangle&=&\sum_{\{\vec\sigma(t)\}}\pi_{\beta(t_i),h(t_i)}
	(\vec\sigma(t_i))\prod_{t=t_i}^{t_f-1}\Big[
	{{\cal Z}_{\beta(t+1),h(t+1)}\pi_{\beta(t+1),h(t+1)}(\vec\sigma(t))\over
	{\cal Z}_{\beta(t),h(t)}\pi_{\beta(t),h(t)}(\vec\sigma(t))}\Big.\nonumber\\
	&&\Big.\hspace{11em}\times W_{\beta(t+1),h(t+1)}(\vec\sigma(t)\rightarrow\vec\sigma(t+1))\Big]
	\end{eqnarray}
All but two partition functions cancel. Using the detailed balance condition (\ref{DetailledBalance}),
one obtains
	\begin{eqnarray}
	\langle X(t_i,t_f)\rangle&=&{{\cal Z}_{\beta(t_f),h(t_f)}\over {\cal Z}_{\beta(t_i),h(t_i)}}
	\sum_{\{\vec\sigma(t)\}}\pi_{\beta(t_i),h(t_i)}(\vec\sigma(t_i))\prod_{t=t_i}^{t_f-1}\Big[
	{\pi_{\beta(t+1),h(t+1)}(\vec\sigma(t+1))\over \pi_{\beta(t),h(t)}(\vec\sigma(t))}\Big.	\\
	&&\Big.\hspace{11em}\times W_{\beta(t+1),h(t+1)}(\vec\sigma(t+1)\rightarrow\vec\sigma(t))\Big]
	\nonumber\end{eqnarray}
Now all but one equilibrium probabilities cancel and one gets
	\begin{equation}
	\langle X(t_i,t_f)\rangle={{\cal Z}_{\beta(t_f),h(t_f)}\over {\cal Z}_{\beta(t_i),h(t_i)}}
	\sum_{\{\vec\sigma(t)\}} \pi_{\beta(t_f),h(t_f)}(\vec\sigma(t_f))\prod_{t=t_i}^{t_f-1}
	W_{\beta(t+1),h(t+1)}(\vec\sigma(t+1)\rightarrow\vec\sigma(t))
	\end{equation}
Finally, the sum over histories of the system is easily shown to be equal to 1 using equation
(\ref{eq2}). Introducing the definition of the free-energy $F(\beta,h)=-\beta^{-1}\ln{\cal Z}_{\beta,h}$,
one obtains the Jarzynski relation
	\begin{equation}
	\langle e^{-\sum_{t=t_i}^{t_f-1} (\Delta_t\beta E(\vec\sigma(t))
	-\Delta_t(\beta h)M(\vec\sigma(t)))}\rangle
	=e^{-[\beta(t_f)F(\beta(t_f),h(t_f))-\beta(t_i)F(\beta(t_i),h(t_i))]}
	\label{Jarzynski}
	\end{equation}
In this paper, we restrict ourselves to the case $h=0$, i.e.
	\begin{equation}
	\langle e^{-\sum_{t=t_i}^{t_f-1} \Delta_t\beta E(\vec\sigma(t))}\rangle
	=e^{-[\beta(t_f)F(\beta(t_f))-\beta(t_i)F(\beta(t_i))]}
	\label{Jarzynskib}
	\end{equation}
\\

The relation (\ref{Jarzynski}) is appropriate for Monte Carlo simulations since the time
is discrete. The demonstration given by Jarzynski in the case of a Markovian process in
continuous time can also be extended to a varying temperature~\cite{Jarzynski97b}.
It leads to
	\begin{equation}
	\langle e^{-\int_{t_i}^{t_f}[\dot\beta U(\vec\sigma(t))-\beta\dot hM(\vec\sigma(t))]dt}
	\rangle=e^{-[\beta(t)F(\beta(t),h(t))]_{t_i}^{t_f}}
	\label{Jarzynski2}
	\end{equation}
where $U(\vec\sigma(t))=E(\vec\sigma(t))-h(t)M(\vec\sigma(t))$ is the total energy of the system
in the state $\vec\sigma(t)$.

\end{document}